   \documentclass[10pt,twoside]{article}
\begin{document}

  \noindent
  \Large
  {\bf Bohmian mechanics in relativistic quantum mechanics,
  quantum field theory and string theory}
  \normalsize
  \vspace*{1cm}

  {\bf Hrvoje Nikoli\'c} 
  
  Theoretical Physics Division, Rudjer Bo\v{s}kovi\'{c} Institute,
  P.O.B. 180, HR-10002 Zagreb, Croatia 

  E-mail: hrvoje@thphys.irb.hr



\begin{abstract}
I present a short overview of my recent achievements 
on the Bohmian interpretation
of relativistic quantum mechanics, quantum field theory
and string theory.
This includes the relativistic-covariant Bohmian equations for
particle trajectories,
the problem of particle creation and destruction,
the Bohmian interpretation of fermionic fields and the
intrinsically Bohmian quantization of fields and strings
based on the De Donder-Weyl covariant canonical formalism.
\end{abstract}

\section{Introduction -- well-established results on the 
Bohmian interpretation}

Let me start with a brief overview of well-established results on
the Bohmian interpretation of quantum mechanics and quantum field theory
\cite{bohm,bohmPR1,bohmPR2,holPR,holbook}.
Consider the Schr\"odinger equation
\begin{equation}
\left[ \frac{-\hbar^2\nabla^2}{2m}+V \right] \psi=i\hbar\partial_t \psi .
\end{equation}
By writing the wave function in the polar form
\begin{equation}
\psi({\bf x},t)=R({\bf x},t)e^{iS({\bf x},t)/\hbar},
\end{equation}
the complex Schr\"odinger equation splits up into two
real equations, the
quantum Hamilton-Jacobi equation
\begin{equation}
\frac{(\nabla S)^2}{2m}+V+Q=-\partial_tS ,
\end{equation}
and the conservation equation
\begin{equation}\label{cons}
\partial_t R^2 + \nabla\left( R^2\frac{\nabla S}{m} \right) =0 ,
\end{equation}
where the quantum potential $Q$ is defined as
\begin{equation}
Q\equiv -\frac{\hbar^2}{2m}\frac{\nabla^2 R}{R} .
\end{equation}
The conservation equation implies that $|\psi|^2$
can be interpreted as a probability density.

The Bohmian interpretation consists in the assumption that
the particle has a deterministic trajectory ${\bf x}(t)$ satisfying
\begin{equation}\label{Ebohm}
\frac{d {\bf x}}{dt}=\frac{\nabla S}{m}.
\end{equation}
This equation has the same form as an analogous equation
in the classical Hamilton-Jacobi theory (see, however, \cite{nikclas}). 
When (\ref{Ebohm}) is combined with the quantum Hamilton-Jacobi equation,
one obtains the quantum Newton equation
\begin{equation}
m\frac{d^2 {\bf x}}{dt^2}=-\nabla(V+Q),
\end{equation}
which has the same form as the classical Newton equation,
except for an additional quantum force generated by the
quantum potential $Q$.
Eq. (\ref{Ebohm}) together with the conservation equation (\ref{cons})
provides that the statistical distribution of particle positions 
always coincides with the quantum distribution $|\psi|^2$, provided
that this distribution coincides with the quantum distribution 
at some arbitary initial time.

To see how the Bohmian interpretation recovers all measurable statistical 
predictions of standard quantum theory, consider a wave function
\begin{equation}
\psi({\bf x})=\sum_a c_a \psi_a({\bf x}) .
\end{equation}
Interaction with the measuring apparatus induces entanglement
resulting in a total wave function of the form
\begin{equation}
\Psi({\bf x},{\bf y}) = \sum_a c_a \psi_a({\bf x}) \phi_a({\bf y}) ,
\end{equation}
where ${\bf y}$ is the degree of freedom of the measuring apparatus.
It is assumed that
the wave functions $\phi_a({\bf y})$ do not overlap, which implies that 
the ${\bf y}$-particle can only be in one channel $\phi_a$. 
Consequently, ${\bf x}(t)$ behaves as if the total wave function were 
$\psi_a \phi_a$, which explains the effective collapse of the
wave function. Thus, in the Bohmian interpretation, there is no need 
for introducing a true collapse.

The above can be easily generalized to the many-particle case.
For a many-particle wave function
$\psi({\bf x}_1, \ldots, {\bf x}_n,t)$ one obtains 
the corresponding many-particle quantum potential 
$Q({\bf x}_1, \ldots, {\bf x}_n,t)$. In general,
this quantum potential is nonlocal, which induces a 
nonlocal instantaneous interaction between particles. 
Thus, the Bohmian interpretation is a
nonlocal hidden-variable interpretation consistent with the Bell theorem. 

There are some attempts to show that, in some cases, 
the predictions of the Bohmian 
interpretation of nonrelativistic quantum mechanics are not 
compatible with standard quantum theory. However, such attempts
seem to be erroneous (see e.g. \cite{st,nikexp} and references therein). 

All the results above can also be generalized to bosonic fields.
Using units $\hbar=1$, quantum field theory for a bosonic field $\phi$
can be described by a functional Schr\"odinger equation
\begin{equation}\label{schfield}
H\left[ -i\frac{\delta}{\delta\phi}, \phi \right] \Psi[\phi;t)=
i\partial_t \Psi[\phi;t) .
\end{equation}
In analogy with (\ref{Ebohm}), the Bohmian interpretation 
assumes that the field has a deterministic dependence on time
determined by the equation
\begin{equation}\label{fieldnc}
\dot{\phi}=\frac{\delta S}{\delta\phi} ,
\end{equation}
which coincides with the analogous classical equation in the 
Hamilton-Jacobi formulation of classical field theory.

The usual formulation of the Bohmian interpretation has several 
open questions, such as the following:
How to make nonlocality consistent with relativity? 
How to introduce the Bohmian interpretation for fermionic fields? 
Should the Bohmian interpretation be applied to particles or fields?
Is there an observable consequence of the Bohmian interpretation? 
Can the Bohmian interpretation be derived (not postulated) from more 
fundamental principles? What about strings? 
In the rest of this paper I present an overview of my
contributions to the attempts of providing answers to all these questions. 

\section{Conservation equation and fermionic fields}

In general, a wave function $\psi(\vec{\varphi},t)$ (where $\vec{\varphi}$
is a many-component continuous degree of freedom) satisfies
a Schr\"odinger equation of the form
$
\hat{H}\psi=i\partial_t \psi 
$,
where $\rho=\psi^*\psi$ is the probability density. 
The average velocity can be calculated as
\begin{equation}
\frac{d \langle \vec{\varphi} \rangle}{dt}=\int d^n\varphi \, \rho \vec{u} ,
\end{equation}
where
\begin{equation}
\vec{u}={\rm Re}\, i\frac{\psi^*[\hat{H},\vec{\varphi}]\psi}
{\psi^*\psi} .
\end{equation}
In general,
\begin{equation}
\partial_t\rho+\vec{\nabla}(\rho\vec{u})=J\neq 0 ,
\end{equation}
where $J$ is some function that, in general, does not need to vanish.
On the other hand, for the consistency of the Bohmian interpretation
written in the form
\begin{equation}
\frac{d\vec{\varphi}}{dt}=\vec{v} ,
\end{equation}
we need a conservation equation of the form
\begin{equation}
\partial_t\rho+\vec{\nabla}(\rho\vec{v})=0 .
\end{equation}
To find $\vec{v}$, we use the ansatz
\begin{equation}
\vec{v}=\vec{u}+\rho^{-1}\vec{\cal E} ,
\end{equation}
where $\vec{\cal E}$ is some new function. From the equations above
one finds that $\vec{\cal E}$ must satisfy
\begin{equation}
\vec{\nabla}\vec{\cal E}=-J .
\end{equation}
Thus, for any $J$, one can find a consistent solution $\vec{\cal E}$
\cite{nikfpl2}.

Now let us apply it to fermionic quantum field theory \cite{nikfpl2}.
Any fermionic state can be written as
\begin{equation}
\Psi^F=\sum_n c_n \Psi^F_n ,
\end{equation}
where
$\Psi^F_n$ is an $n$-particle state.
For each fermionic state $\Psi^F_n$ there is a corresponding bosonic state
$\Psi^B_n$
with the same number of particles having the same momenta. This allows us
to introduce a map
\begin{equation}
\Psi^F_n \rightarrow \Psi^B_n[\phi] .
\end{equation}
(The reverse is not true, because there are more bosonic states than fermionic
ones.) 
This provides a bosonic representation of fermionic states. 
Consequently, the Bohmian interpretation is possible in a similar
way as that for bosonic fields, with the 
conservation equation provided as above.  

\section{Relativistic quantum mechanics}

To make the Bohmian interpretation of particles compatible 
with the Bohmian interpretation of fields,
I propose that both fields and particle positions are fundamental 
entities \cite{nikfpl1,nikfpl2}.
Consider the field operator satisfying the Klein-Gordon equation
$
(\partial^{\mu}\partial_{\mu}+m^2)\hat{\phi}(x)=0 
$,
where $\hat{\phi}$ is a hermitian (uncharged) field. The 
$n$-particle wave function is
\begin{equation}\label{e21}
\psi(x_1,\ldots ,x_n)=(n!)^{-1/2}S_{\{ x_a\} }
\langle 0|\hat{\phi}(x_1)\cdots\hat{\phi}(x_n)|n\rangle ,
\end{equation}
where $S_{\{ x_a\} }$ denotes the symmetric ordering. From this
$n$-particle wave function one can construct
$n$ conserved particle currents (one for each $a$):
\begin{equation}
j^{\mu}_a=i\psi^* \!\stackrel{\leftrightarrow\;}{\partial^{\mu}_a}\! \psi .
\end{equation}
(For more details on particle currents in quantum field theory, see also
\cite{nikcurr1,nikcurr2,nikcurr3}.)
The $n$-particle wave function satifies the $n$-particle Klein-Gordon 
equation, which leads to the
relativistic quantum potential
\begin{equation}
Q=\frac{1}{2m}\frac{\sum_a\partial_a^{\mu}\partial_{a\mu}R}{R} .
\end{equation}
Here
$Q(x_1,\ldots ,x_n)$ is nonlocal, but relativistic invariant!
The Bohmian interpretation consists in the assumption that 
particle trajectories satisfy
\begin{equation}\label{bohmreln}
\frac{dx_a^{\mu}}{ds} = -\frac{1}{m}\partial_a^{\mu}S 
= \frac{j_a^{\mu}}{2m\psi^*\psi} .
\end{equation}
The trajectories in spacetime do not depend on the choice of the 
auxiliary parameter $s$, which can be seen by writing
(\ref{bohmreln}) as 
\begin{equation}
\frac{dx_a^{\mu}}{dx_b^{\nu}} = \frac{j_a^{\mu}}{j_b^{\nu}} ,
\end{equation}
which eliminates $s$. For a more detailed discussion 
of the covariance of these Bohmian equations of motion, see
\cite{nikproclosinj}. Note also that these trajectories 
may correspond to superluminal motions and motions backwards
in time, but that this does not lead to causal paradoxes
(see also \cite{nikcaus}).
To find a measurable consequence of this relativistic 
Bohmian equation of motion,
assume that the interaction with the apparatus that measures particle 
positions starts at some particular time $t_1$. 
One finds that \cite{nikfpl3}
\begin{equation}
\rho({\bf x},t_1)=\left\{  
\begin{array}{ll}
j_0({\bf x},t_1) & \mbox{on $\Sigma'$} , \\
0 & \mbox{on $\Sigma^+\cup\Sigma^-$} ,
\end{array}   
\right. 
\end{equation}
where $\Sigma^-$ is the part of the $t_1$-hypersurface at which 
$j_0<0$, $\Sigma^+$ is the part of the $t_1$-hypersurface
which is connected with $\Sigma^-$ by Bohmian trajectories 
for which $t<t_1$, and $\Sigma'$ represents all other points of the
$t_1$-hypersurface. We emphasize that
this measurable result cannot be obtained without 
calculating the trajectories. We also emphasize that
there is no ``standard" prediction for the case $j_0<0$
\cite{nikno1,nikmyth}, 
while the Bohmian interpretation provides a prediction as 
above (see also \cite{nikfol}).

\section{Particle creation and destruction}

What happens with the Bohmian trajectories when particles are
created or destructed? 
A possible answer is that nothing dramatic happens, 
in the sense that there are no sudden creation or destruction of
particle trajectories.   
Instead, the trajectories exist all the time, 
but their effectivities (to be defined below) 
change continuously with time \cite{nikfpl1,nikfpl2}.
Consider a QFT state
\begin{equation}
\Psi=\sum_n \tilde{\Psi}_n ,
\end{equation}
where the tilde on $\tilde{\Psi}_n$ denotes that the norm of this
$n$-particle state 
can be smaller than $1$.
From (\ref{e21}), we see that an
$n$-particle wave function is essentially
\begin{equation}
\psi_n \propto \langle 0|\hat{\phi}(x_1) \cdots 
\hat{\phi}(x_n)|\Psi\rangle .
\end{equation}
The important fact is that the
Bohmian trajectories do not depend on the norm of $\psi_n$. 
The effectivity is defined as
\begin{equation}
e_n[\phi;t) = \displaystyle\frac{ |\tilde{\Psi}_n[\phi;t)|^2 }
{ \displaystyle\sum_{n'} |\tilde{\Psi}_{n'}[\phi;t)|^2 } ,  
\end{equation}
which, for a definite Bohmian value of $\phi$,
measures the relative contribution of the $n$-particle sector.
When $n$ is measured, then the Bohmian evolution of $\phi$ explains the
effective collapse $\Psi\rightarrow\Psi_n$. 
In this case, $e_n=1$ for this $n$ and $e_{n'}=0$ for all other $n'$,
i.e. only the particles of the $n$-particle sector are effective.

\section{Bohmian mechanics from covariant quantization of fields
and strings}
  
The Bohmian equation of motion (\ref{fieldnc}) is not 
relativistic covariant.
This is related to the fact that the functional Schr\"odinger 
equation (\ref{schfield}) is also not covariant.
To solve this problem, one possibility is to base 
the Bohmian interpretation of fields on the many-fingered time
Tomonaga-Schwinger equation \cite{nikpla,nikmft}, but the 
existing results do not seem to be completely satisfying \cite{privcom},
except for theories that satisfy a Hamiltonian constraint \cite{nikmft}, 
such as theories that contain quantum gravity.
  
Another strategy is to modify the quantization itself to make it 
manifestly relativistic covariant in a different way 
\cite{nikepjc1,nikessay}.
Instead of the noncovariant Hamilton formalism based on the Hamiltonian
$
{\cal H}=\pi\partial_0\phi-{\cal L} 
$
with the canonical momentum
$
\pi=\partial {\cal L} / \partial(\partial_0\phi) 
$,
I use De Donder-Weyl covariant Hamilton formalism based on the 
covariant Hamiltonian
\begin{equation}
{\cal H}^{\rm DW}=\pi^{\mu}\partial_{\mu}\phi-{\cal L} ,
\end{equation}
where the covariant canonical momentum is a vector
\begin{equation}
\pi^{\mu}=\frac{\partial {\cal L}}{\partial(\partial_{\mu}\phi)} .  
\end{equation}
This leads to the 
covariant Hamilton-Jacobi equation
\begin{equation}
{\cal H}^{\rm DW} \left( \frac{\partial S^{\alpha}}{\partial\phi}, \phi
\right) +\partial_{\mu}S^{\mu}=0 ,
\end{equation}
and the
covariant equation of motion
\begin{equation}\label{covem}
\partial^{\mu}\phi=\frac{\partial S^{\mu}}{\partial\phi} .
\end{equation}
To derive the noncovariant Hamilton-Jacobi equation with
${\cal H}^{\rm DW}\rightarrow {\cal H}$, it is necessary \cite{nikepjc1} to use 
the spacial part 
$
\partial^i\phi=\partial S^i / \partial\phi 
$
of (\ref{covem}).
The covariance then implies that the time part
$
\partial^0\phi=\partial S^0 / \partial\phi
$
must also be valid. In other words, the determinism of classical
mechanics is derived from the requirement of covariance!

To quantize the theory, one modifies the covariant Hamilton-Jacobi 
equation by adding the quantum potential to the classical
Hamiltonian:
\begin{equation}
{\cal H}^{\rm DW}_Q={\cal H}^{\rm DW}+Q .
\end{equation}
Here $Q$ is the same as that for the functional Schr\"odinger equation, 
but is written in a covariant form \cite{nikepjc1}.
In the same way as for classical fields, to derive the Schr\"odinger equation,
the spacial part of (\ref{covem}) must be valid.
Covariance then implies the time part of (\ref{covem}). 
This is a derivation of the Bohmian equation of motion from the 
requirement of covariance!

The idea above can also be applied to strings. This is because
strings can be viewed as fields in 2 dimensions. 
Thus, in the same way, the world-sheet covariance implies 
the Bohmian equation of motion for strings \cite{nikepjc2}.
In the Bohmian interpretation,
the string always has a well-defined shape
$X^{\mu}(\sigma,\tau)$, even when it is not measured. 
On the other hand, observable properties of strings, 
such as the mass-spectrum, obey T-duality (see e.g. \cite{zwie}),
which is a symmetry under
the change of the compactification radius $R$ with the dual
radius $\alpha'/R$ (where $\sqrt{\alpha'}$ is the fundamental
length scale of string theory). It is widely believed that
T-duality is a fundamental symmetry of string theory that makes 
the theory nonlocal at distances smaller than $\sqrt{\alpha'}$. 
However, the Bohmian interpretation breaks T-duality at the 
fundamental level of hidden variables \cite{niktdual}, 
which makes the role of T-duality less fundamental.

\section{Conclusion}

Bohmian mechanics is more ambitious (and thus more complicated)
than standard quantum theory. The goals of Bohmian mechanics are 
(i) to recover the predictions of standard quantum theory 
and (ii)
to explain (like classical mechanics and unlike standard quantum theory) 
what is going on when measurements are {\em not} performed.
The existing results indicate that it is possible to achieve this for
all quantum phenomena. 

\section*{Acknowledgments}

This work was supported by the Ministry of Science and Technology of the
Republic of Croatia.
\\


\begin{thebibliography}{99}

\bibitem{bohm}
Bohm D 1952 {\it Phys.~Rev.}~{\bf 85} 166, 180 
\bibitem{bohmPR1}
Bohm D and Hiley B J 1987
{\it Phys.~Rep.}~{\bf 144} 323 
\bibitem{bohmPR2}
Bohm D, Hiley B J and Kaloyerou P N 1987
{\it Phys.~Rep.}~{\bf 144} 349 
\bibitem{holPR}
Holland P R 1993 {\it Phys.~Rep.}~{\bf 224} 95 
\bibitem{holbook}
Holland P R 1993 {\it The Quantum Theory of Motion}
(Cambridge: Cambridge University Press)

\bibitem{nikclas}
Nikoli\'c H 2006 {\it Found. Phys. Lett.} (in press)
({\it Preprint} quant-ph/0505143)

\bibitem{st}
Struyve W, De Baere W, De Neve J and De Weirdt S 2003 {\it J. Phys.} A 
{\bf 36} 1525

\bibitem{nikexp}
Nikoli\'c H 2003 {\it Preprint} quant-ph/0305131

\bibitem{nikfpl2}
Nikoli\'c H 2005 {\it Found. Phys. Lett.} {\bf 18} 123

\bibitem{nikfpl1}
Nikoli\'c H 2004 {\it Found. Phys. Lett.} {\bf 17} 363

\bibitem{nikcurr1}
Nikoli\'c H 2002
{\it Phys. Lett.} B {\bf 527} 119

\bibitem{nikcurr2}
Nikoli\'c H 2003
{\it Int. J. Mod. Phys.} D {\bf 12} 407

\bibitem{nikcurr3}
Nikoli\'c H 2005
{\it Gen. Rel. Grav.} {\bf 37} 297

\bibitem{nikproclosinj}
Nikoli\'c H 2006 
{\it AIP Conf. Proc.} {\bf 844} 272
({\it Preprint} quant-ph/0512065)

\bibitem{nikcaus}
Nikoli\'c H 2006 {\it Found. Phys. Lett.} {\bf 19} 259

\bibitem{nikfpl3}
Nikoli\'c H 2005 {\it Found. Phys. Lett.} {\bf 18} 549

\bibitem{nikno1}
Nikoli\'c H 2003 {\it Preprint} quant-ph/0307179

\bibitem{nikmyth}
Nikoli\'c H 2006 {\it Preprint} quant-ph/0609163 

\bibitem{nikfol}
Nikoli\'c H 2006 {\it Preprint} quant-ph/0602024

\bibitem{nikpla}
Nikoli\'c H 2006 {\it Phys. Lett.} A {\bf 348} 166

\bibitem{nikmft}
Nikoli\'c H 2006 {\it Preprint} quant-ph/0603207

\bibitem{privcom}
Goldstein S and Tumulka R 2006 private communication

\bibitem{nikepjc1}
Nikoli\'c H 2005 {\it Eur. Phys. J.} C {\bf 42} 365

\bibitem{nikessay}
Nikoli\'c H 2006  
{\it Honorable Mention of the Gravity Research Foundation 
2006 Essay Competition}
({\it Preprint} hep-th/0601027)

\bibitem{nikepjc2}
Nikoli\'c H 2006 {\it Eur. Phys. J.} C {\bf 47} 525

\bibitem{zwie}
Zwiebach B 2004 {\it A First Course in String Theory} 
(Cambridge: Cambridge University Press)

\bibitem{niktdual}
Nikoli\'c H 2006 {\it Preprint} hep-th/0605250

\end{thebibliography}
\end{document}